\documentclass[multphys,vecphys]{svmult}

\usepackage{makeidx}         
\usepackage{graphicx}        
\usepackage{multicol}        
\usepackage[bottom]{footmisc}

\makeindex             

\begin{document}

\title*{Jet deceleration: the case of PKS 1136-135}
\author{F. Tavecchio\inst{1}, L. Maraschi\inst{1}\and
R.M. Sambruna\inst{2}}

\institute{INAF-OAB, via Brera 28, 20121 Milano, Italy
\and NASA/GSFC, Code 661, Greenbelt, MD, 20771, USA}
%
%
\maketitle

\section{Introduction}
\label{sec:1}

Despite decades of intense efforts, the present knowledge of the
physical processes acting in relativistic jets is still rather poor
and basic questions are awaiting answers (e.g. Blandford 2001). Among
these problems, one of the most fundamental concerns the speed of the
flow and the processes leading to deceleration. The present evidence
suggests that FRI jets decelerate, becoming trans-relativistic, quite
early, within few kiloparsecs (e.g. Bridle \& Perley 1984), while the
situation for FRII jets appears more ambiguous. The interpretation of
multiwavelength observations of extended jets in QSOs points toward
highly relativistic speeds ($\Gamma \sim 10$) even at very large
scales ($\sim $ 100 kpc; Tavecchio et al. 2000, Celotti et al. 2001;
see Stawarz et al. 2004 and Atoyan \& Dermer 2004 for some criticisms
to the model
). In the same sources, recent multiwavelength observations suggest that
also these jets undergo deceleration close to their terminal hot
spots. This (model-dependent) conclusion is based on the observed
increase of the radio to X-ray flux along the jet, which, interpreted in
the framework of the synchrotron-IC/CMB emission model, uniquely implies
deceleration.

The possibility of ``observing'' the gradual slowing-down of a jet
could in principle provide precious information on the physical
processes at work. This approach was successfully developed in great
detail for a few FRI jets where the morphology could be well studied
thanks to the large angular scale (e.g. Laing \& Bridle 2002).  Here we
report on the analysis performed on the FRII jet of the quasar PKS
1136-135, for which the excellent data (Sambruna et al. 2006), allow
us to apply a similar (but less detailed) approach. More details can
be found in Tavecchio et al. (2006).

\section{Modelling deceleration}

The profiles of the interesting physical quantities of the jet (the bulk
Lorentz factor, $\Gamma$, the intensity of the magnetic field, $B$, and
the density of the non-thermal electrons, $K$) can be derived by applying
the IC/CMB emission model to the measured radio and X-ray fluxes at
different emission knots along the jet. The derived values for different
regions in the jet are shown in Fig. 1. The errorbars take into account
as much as possible all the uncertainties (associated to the measurements
and the modelling) affecting the derivation of the parameters. Clearly,
the Lorentz factor appears to decrease along the jet, going from $\Gamma
\sim 6$ at B to $\Gamma \sim 2.5$ at F. At the same time, the inferred
magnetic field and the particle density increase, as expected in the case
of deceleration (Georganopoulos \& Kazanas 2004).

\begin{figure}
\begin{tabular}{cc}
\hspace{-2.cm}
\includegraphics[height=8cm]{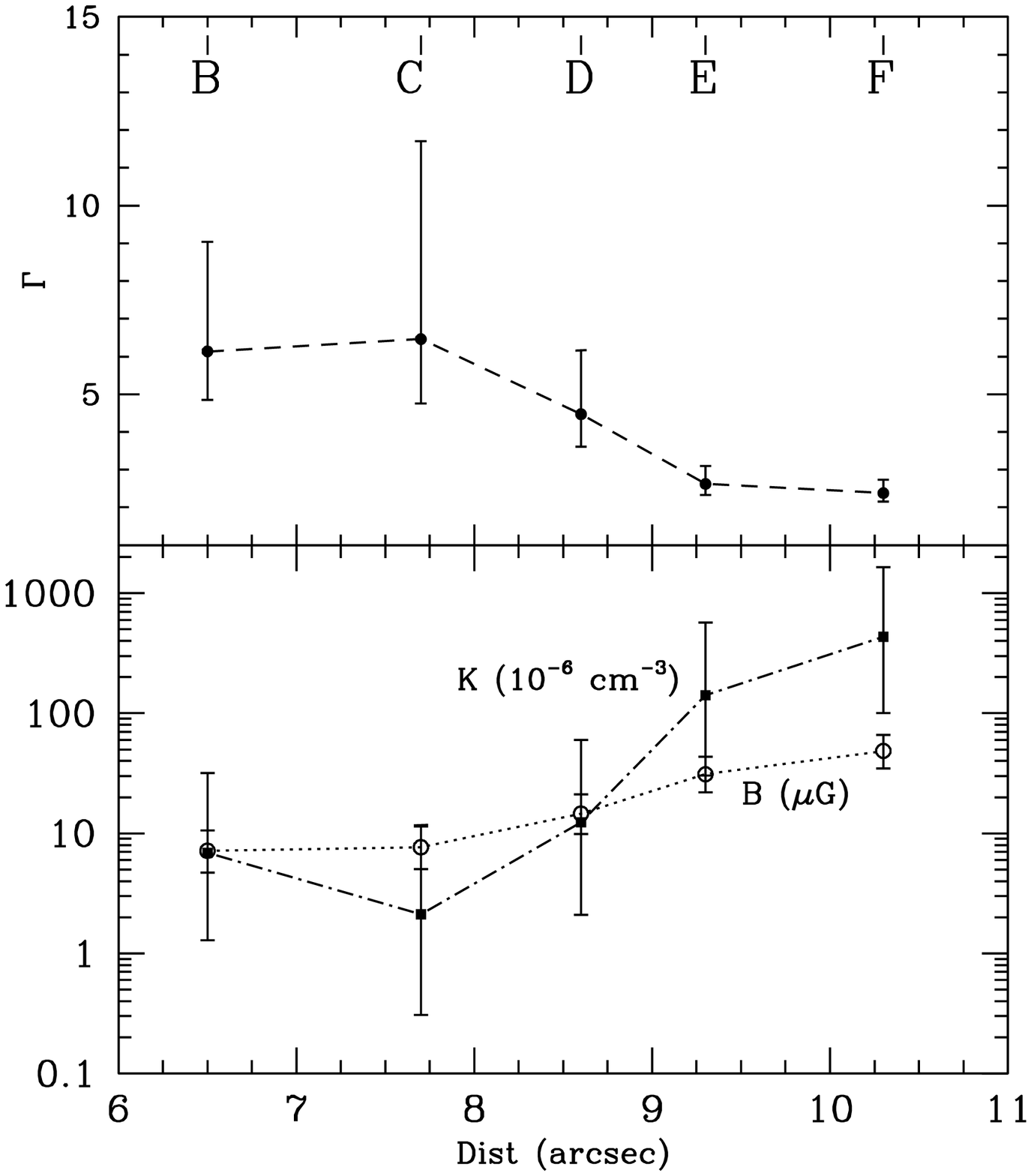}
&
\hspace{-1.2cm}
\includegraphics[height=8cm]{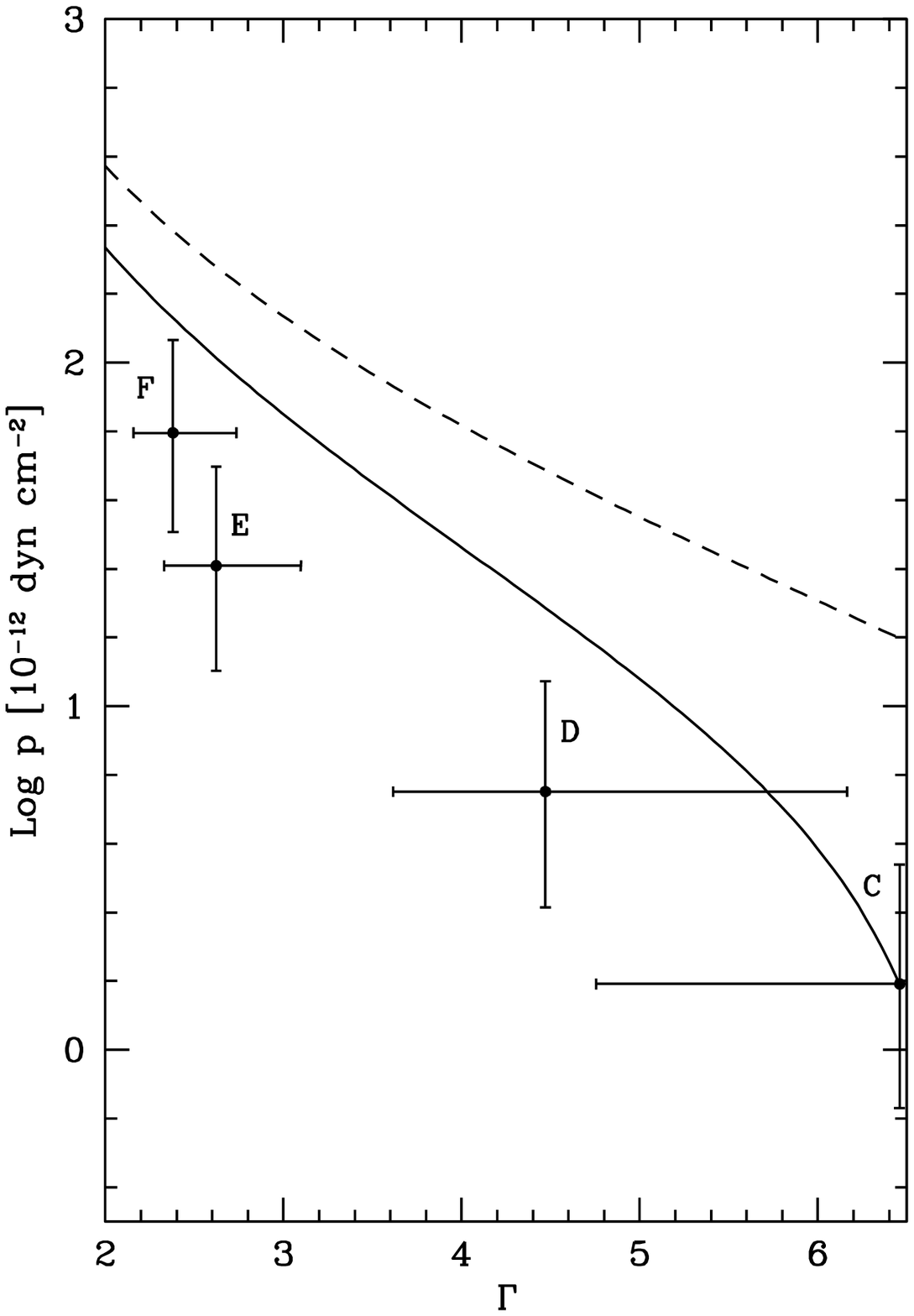}
\end{tabular}
\caption{\footnotesize {\it Left Panel:} profiles of the relevant
quantities ($\Gamma$, top panel, $B$ and $K$, lower panel) for regions
B--F of the jet of PKS 1136-135 estimated from the radiative
model. {\it Right Panel:} the pressure inside the jet as a function of
the Lorentz factor of the jet, calculated with the momentum and energy
flux conservation laws (Bicknell 1994), assuming the initial
conditions inferred for knot C. The solid line refers to the case in
which the initial pressure of protons is negligible, while the dashed
line is calculated assuming an initial pressure in protons ten times
that supported by the non-thermal component (relativistic electrons
and magnetic field). Crosses indicate the value of the non-thermal
pressure (provided by magnetic field and non-thermal electrons)
calculated through the modelling of the observed emission (both plots
are adapted from Tavecchio et al. 2006).}
\label{fig:1}
\end{figure}

In the case in which the jet inertia is dominated by protons (as
supported by several indications, e.g. Maraschi \& Tavecchio 2003), we
explored the possibility that entrainment of external gas is effective in
decelerating the jet. Basically, deceleration through entrainment can
be understood to happen through a continuous series of inelastic
collisions between the moving plasma and the external gas at rest. As
a result of the collision, part of the kinetic energy is dissipated
and converted into internal energy of the jet, thus increasing the
internal pressure.  We applied the hydrodynamical treatment, based on
the use of energy and momentum conservation, developed by Bicknell
(1994) to describe the deceleration of the jet of 1136-135 in order to
discuss the plausibility of the entrainment mechanism for this
particular case. The predicted run of the pressure is reported in
Fig. 1.

\section{Discussion}

The proposed scenario can explain in a plausible way the deceleration
inferred for the jet of 1136-135 (and possibly in the other sources
showing the same radio-to-X-rays increasing trend). A more detailed
understanding of the processes at work and the comparison with the
observed properties of jets necessarily involves several still
poorly-known physical issues. An important feature of the
entrainment-induced deceleration is that the jet starts to slow down
significantly when the mass of collected gas is of the order of 1/$\Gamma
$ of the mass transported by the jet. In this framework, jets
characterized by different mass fluxes will experience different
behaviours. Large mass fluxes will assure that, under the same conditions
of external gas density and entrainment rate, the jet will reach its
hotspot almost unperturbed.  On the other hand, jets with a small mass
flux will be decelerated soon. It is tempting to further speculate along
these lines, associating the FRII morphology to jets with large mass flux
and FRI objects to jets characterized by small mass fluxes.

%
%

%
%
%
%

%
%



\printindex
\end{document}